\documentclass[aps,floatfix,twocolumn,final,prl]{revtex4}

\usepackage{epsfig}
\usepackage{graphicx}
\usepackage[ansinew]{inputenc}
\usepackage{array}
\usepackage{hyperref}
\usepackage{tabularx}

\begin{document}
%\preprint{APS/123-QED}
\title{Revisiting the Possible 4$f^{7}$ 5$d^{1}$ Ground State of Gd Impurities in SmB$_{6}$ by Electron Spin Resonance}% Force line breaks with \\

\author{J. C. Souza$^{1,2}$, P. F. S. Rosa$^{3}$, U. Burkhardt$^{2}$, M. K\"onig$^{2}$, Z. Fisk$^{4}$, P. G. Pagliuso$^{1}$, S. Wirth$^{2}$ and J. Sichelschmidt$^{2}$}

\affiliation{$^{1}$Instituto de F\'isica \lq\lq Gleb Wataghin\rq\rq,
UNICAMP, 13083-859, Campinas, SP, Brazil\\
$^{2}$Max Planck Institute for Chemical Physics of Solids, D-01187 Dresden, Germany\\
$^{3}$Los Alamos National Laboratory, Los Alamos, New Mexico 87545, USA\\
$^{4}$Department of Physics and Astronomy, University of California, Irvine, California 92697,USA}

\email{jcsouza@ifi.unicamp.br}

\date{September 17, 2019}

%  %\homepage{http://www.Second.institution.edu/~Charlie.Author}
% %\affiliation{

\begin{abstract}

The search for topological states in strongly correlated electron systems has renewed the interest in the Kondo insulator SmB$_{6}$. One of the most intriguing previous results was an anomalous electron spin resonance spectrum in Gd-doped SmB$_{6}$. This spectrum was attributed to Gd$^{2+}$ ions because it could be very well decribed by a model considering a change in the valence from Gd$^{3+}$ to Gd$^{2+}$, a dynamic Jahn-Teller effect and a 4$f^{7}$ 5$d^{1}$ ground state in the Hamiltonian. In our work, we have revisited this scenario using electron spin resonance and energy dispersive X-ray spectroscopy measurements. Our results suggest that the resonance is produced by Gd$^{2+}$ ions; however the resonance stems from an extrinsic oxide impurity phase that lies on the surface of the crystal.
\end{abstract}

%\keywords{electron spin resonance, semiconductors, heavy-fermion materials}

%\pacs{76.30.-v, 71.20.Lp}% PACS, the Physics and Astronomy maybe 75.40.Gb dunamic, ato magnetic critical points
\maketitle

\section{Introduction}

The mixed-valence ground state, the nature of the hybridization gap and the saturation of the resistivity below $\approx$ 4 K have been puzzling features of SmB$_{6}$ for a long time \cite{dzero2016topological}. Since the theoretical prediction of a topological Kondo insulating phase in SmB$_{6}$ \cite{dzero2012theory}, this system has renewed the interest of the community to understand all of the puzzling open questions. One important aspect we emphasize here is the necessity of diagnosing extrinsic effects in this system \cite{thomas2018quantum}.

One of the remaining questions is the possible existence of a formal Gd$^{2+}$ ground state (4$f^{7}$ 5$d^{1}$) of Gd-doped SmB$_{6}$. The extra electron brought by Gd$^{2+}$ to the lattice may be bound in a singlet state which should be relevant for understanding the low-temperature Kondo physics \cite{fisk1996kondo,fisk1996heavy}. Furthermore, a putative existence of a Gd$^{2+}$ state may be relevant in interpreting the suppression of surface states in Gd-doped SmB$_{6}$ \cite{jiao2018magnetic}. Clearly, the role of magnetic impurities on topological and Kondo insulators is essential to understand their fundamental mechanism.

G. Wiese, H. Sch\"affer and B. Elschner reported an electron spin resonance (ESR) study \cite{wiese1990possible} in which a Hamiltonian including a 4$f^{7}$ 5$d^{1}$ ground state and a dynamic Jahn-Teller effect was employed. This was investigated in great detail within a Diploma thesis \cite{wiese1986elektron}, in which ESR measurements at 9.25 GHz (X-band) and various theoretical models were also reported.
%G. Wiese reported in his Diploma \cite{wiese1986elektron} also measurements done with a microwave frequency $\nu$ = 9.5 GHz (X-band) at 4.2 K and discussed various theoretical models. 
Wiese $\textit{et al.}$ measured samples of Sm$_{1-x}$Gd$_{x}$B$_{6}$ in the range of 0.0002 $\leq$ x $\leq$ 0.003 that were grown by an aluminum flux technique \cite{gurin1979}. The authors claimed that the anomalous spectra were observed in all concentrations, but they mainly focused on the analysis of x = 0.003 due to a better signal-to-noise ratio. Furthermore, floating zone grown crystals did not present this supposedly anomalous Gd$^{2+}$ ESR spectra, which was interpreted as an indication that only high quality samples would show such effect.

Aware of the G. Wiese et al. reports \cite{wiese1986elektron,wiese1990possible}, in this work we have measured ESR in Al-flux grown Sm$_{1-x}$Gd$_{x}$B$_{6}$ samples in the range of 0.0002 $\leq$ x $\leq$ 0.004. We were capable of reproducing the X-band measurements at low temperatures reported previously \cite{wiese1986elektron} with the external magnetic field applied along the [100] direction. It is important to note that the Gd ESR anomalous spectrum was also sample dependent in our measurements. 

In order to clarify whether the ESR Gd$^{2+}$ spectra were intrinsic to the system, we performed a detailed study on the surface of the crystals that had the most pronounced signal using energy dispersive X-ray spectroscopy (EDX). We found a clear white region where the surface had an oxide which contains aluminum. Because it was challenging to polish the white region away due to the size of the crystals, we separated a crystal in several pieces and compared the intensity of the ESR Gd$^{2+}$ spectra between two distinct pieces. Remarkably, we observed a significant decrease in the intensity of the Gd anomalous signal in the sample that had a drastically reduced white region, which strongly indicates an extrinsic origin of such effect. This decrease scales particularly well with the area which contains this oxide. 

Recent reports \cite{completing2013macdonald,shallow2018kitaura} demonstrate the possibility of the occurrence of a Gd$^{2+}$ ion with a 4$f^{7}$ 5$d^{1}$ ground state in oxides, which agrees with our observations. Our results point to the possibility that this anomalous Gd$^{2+}$ ESR spectra is coming from an impurity on the surface of the material.

\section{Experiment}

Single crystalline samples of Sm$_{1-x}$Gd$_{x}$B$_{6}$ were Al-flux grown as described elsewhere \cite{eo2019transport}. Elemental analyses were performed using energy dispersive X-ray spectroscopy (EDX) with two different commercial equipment. We performed measurements on etched and non-etched samples to investigate whether the signal was due to some impurity on the surface that could be etched away. The etching process was done using a dilute mixture of hydrochloric and nitric acids (aqua regia), and the crystals were etched for two minutes in each session. ESR measurements were performed in a X-band (9.4 GHz) spectrometer equipped with a goniometer and a He-flow cryostat, in the temperature range of 2.5 K $\leq$ $T$ $\leq$ 20 K. The studied samples ranged in mass from 0.4 mg to 1.5 mg. We performed our experiments on 10 different crystals from seven different batches. 

\section{Results and Discussion}

\begin{figure}[!ht]
\centering
\includegraphics[width=0.95\columnwidth]{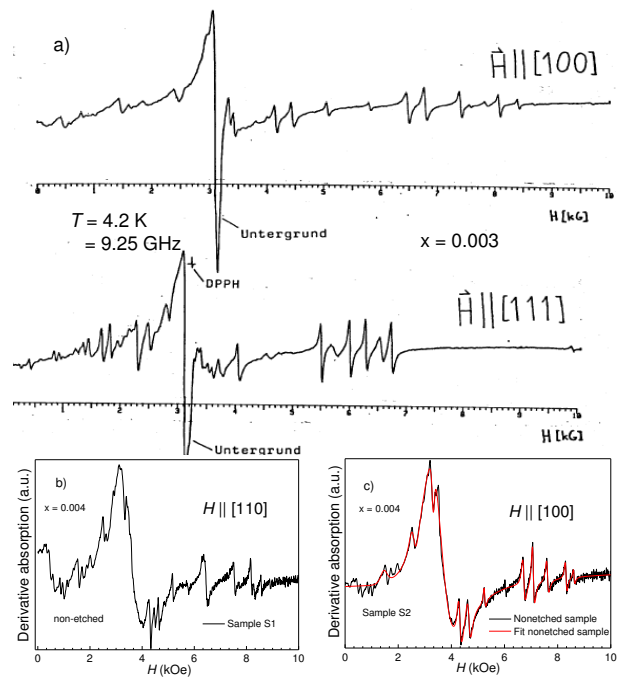}
\caption{a) X-band Gd ESR spectra of Sm$_{1-x}$Gd$_{x}$B$_{6}$ (x=0.003) along the [100] and [111] direction reported by G. Wiese in his Diploma \cite{wiese1986elektron}. b) Samples S1 and c) S2 Gd X-band ESR spectrum at 4.5 K with x = 0.004. The Gd ESR spectrum of S1 was taken with the external magnetic field parallel to the [110] direction and S2 in the [100] direction. The signal at low fields in ($H$ $\leq$ 1 kOe) in b) is due to a contribution from the cavity.}
\label{Fig1}
\end{figure}

Figure \ref{Fig1} a) displays the ESR spectra of Gd in SmB$_{6}$ at 4.2~K with the external magnetic field applied along the [100] and [111] directions obtained by G. Wiese and reported in his Diploma \cite{wiese1986elektron}. Figures. \ref{Fig1} b) and c) show the Gd ESR spectra we obtained for two different crystals (samples S1 and S2) with $x$ $\approx$ 0.004 at 4.5 K and within the same field range $H \le 10$~kOe. Both measurements were performed using crystals that were not etched. Sample S2 was aligned with the magnetic field parallel to the [100] direction, whereas S1 was aligned with the magnetic field parallel to the [110] direction. It is evident from our Gd ESR spectra that we could reproduce the same results obtained by G. Wiese et al. \cite{wiese1986elektron,wiese1990possible}.

We also performed ESR experiments as a function of temperature and field orientation (not shown). The ESR Gd anomalous spectra has a Curie-Weiss behavior and the intensity for $T$ $\geq$ 6 K is reduced drastically, which confirms previous reports \cite{wiese1986elektron,wiese1990possible}. The angle dependence of the anomalous ESR spectra is extremely anisotropic, with no clear cubic symmetry. This is another hint that this anomalous spectra could be connected with extrinsic phases, and not with the Gd ESR in SmB$_{6}$. 

The fine structure for SmB$_{6}$ crystals with 200 ppm and 400 ppm Gd shows a clear cubic symmetry \cite{souza2018esr} in angle-dependent ESR experiments, which confirms that the probe is in the cubic SmB$_{6}$ matrix. In the case of the anomalous Gd spectra, even variations of two degrees had a major impact on the spectrum, with lines vanishing and new ones appearing (not shown). Moreover, the appearance of the anomalous Gd spectra was sample dependent. G. Wiese pointed out \cite{wiese1990possible} that SmB$_{6}$ Gd-doped floating zone crystals grown by Kunii et al. \cite{kunii1985} did not show the anomalous Gd spectrum and explained this result in terms of the quality of the crystals used.

\begin{figure}[!ht]
\centering
\includegraphics[width=0.95\columnwidth]{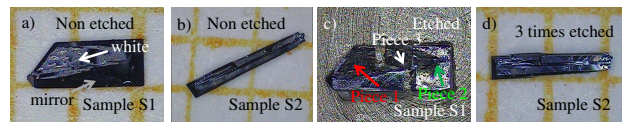}
\caption{Optical microscopy photos of SmB$_{6}$ Gd-doped samples a) S1 non etched, b) S2 non etched, c) S1 etched and broken in three pieces and d) S2 three times etched. All samples had a characteristic white region, as discussed in the text, which differs from pristine SmB$_{6}$.}
\label{Fig2}
\end{figure}

Therefore, we searched for any kind of impurities using optical microscopy. Figure \ref{Fig2} a) shows an image taken by optical microscopy of the Gd-doped SmB$_{6}$ sample S1. We observe a clear difference: there is one region with a more smooth, mirror-like surface, and a white pocket region with a more pronounced roughness. The pattern was also observed for sample S2, as shown in Fig. \ref{Fig2} b). This white region could not be etched away, even by etching the samples for one session, as shown in Figure \ref{Fig2} c), or three sessions, as shown in Fig. \ref{Fig2} d), increasing the aqua regia concentration after each step. This is a clear indication that such regions could be connected with an oxide extrinsic phase.

\begin{figure}[!ht]
\centering
\includegraphics[width=0.95\columnwidth]{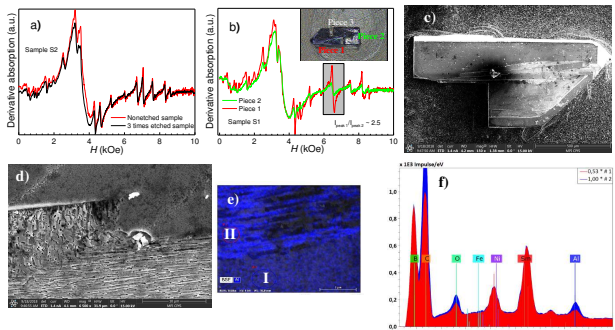}
\caption{Normalized Gd ESR spectra at $T=4.5$K for Gd-doped SmB$_{6}$ samples a) S2 before and after etching and b) two different pieces from S1. c) and d) EDX images of piece 1 from sample S1. The crystals were not etched or polished after the ESR measurements. e) Two distinct points of EDX analysis done in sample S1. The region I corresponds to the red spectrum and the region II to the blue spectrum in f). The Ni and Fe peaks are due to contamination while preparing the sample to perform EDX measurements.}
\label{Fig3}
\end{figure}

After etching the crystals, we performed ESR experiments again in the Gd-doped SmB$_{6}$ samples S1 and S2. As it is shown for S2 in Fig \ref{Fig3} a), when we compare the normalized (by the Gd$^{3+}$ resonance intensity) Gd anomalous ESR spectra, we do not observe a clear change in the Gd anomalous spectra. 

As the Gd-doped SmB$_{6}$ crystals are thin and small it is challenging to polish the samples and obtain a ESR signal due to the intensity of the signal. Furthermore, the white region could be found on different facets of the same crystal, which dashes any hope to improve a crystal surface treatment by  just polishing the samples. In order to avoid such challenge, we intentionally broke a Gd-doped SmB$_{6}$ sample in three different pieces, as it is shown in Figure \ref{Fig2} c), and performed ESR experiments on piece 1 and piece 2 with different surface coverage by white regions. Piece 1 had a larger surface area covered by this rough region compared with piece 2 and the ratio of the areas with white region from piece 1 to piece 2 was $\approx 3.0$, whereas the ratio of the total area was $\approx$ 1.3 and the mass ratio was $\approx$ 1.2.

Figure \ref{Fig3} b) shows the Gd ESR spectra, normalized by the Gd$^{3+}$ intensity from two different pieces - see inset. It is important to emphasize that all the pieces together reproduced the results obtained by G. Wiese et al. \cite{wiese1986elektron,wiese1990possible}, as shown in Fig. \ref{Fig1}. If the Gd anomalous spectrum were coming from the bulk of the crystal, it would be expected that the comparison of the normalized Gd ESR spectra would be very similar, which is not the case. There is a clear reduction in the intensity of the Gd anomalous spectra as measured on piece 2, which is a very convincing hint that the Gd anomalous spectra is not related with the bulk of the Gd-doped SmB$_{6}$ crystal.

By choosing an isolated peak, which is featured in the gray region of Fig. \ref{Fig3} b), we calculated the ratio of the intensity of the same resonance peak for pieces 1 and 2 of sample S1. The ratio was $\approx$ 2.5, which is closer to the ratio of the white regions of these two pieces than to the ratio of the total area or mass. All these results are in agreement with our scenario that such Gd anomalous signal is closely connected with this rough, white surface.

In order to investigate the origin of this white surface, we performed EDX experiments on piece 1 of the sample S1 of Gd-doped SmB$_{6}$ after the ESR measurements reported in Fig. \ref{Fig3} a) and b) to investigate the crystal surface. From a morphological point of view, as shown in Figs. \ref{Fig3} c) and d), the two regions are easily distinguishable: while the gray region is much more flat and smooth, the white region is more rough and irregular. The amplified image shown in Fig. \ref{Fig3} d) clarifies even more the difference between these two regions in the Gd-doped SmB$_{6}$.

Finally, we tried to differentiate, in a semi-quantitative approach, the elements from these two regions of the same crystal. We analyzed the smooth surface (region I) of piece 1 from the sample S1 of Gd-doped SmB$_{6}$, shown in figure \ref{Fig3} e), and compared it with the white surface (region II). Figure \ref{Fig3} f) shows the comparison between the obtained spectra, where the red spectrum was obtained for the smooth surface and the blue for the rough surface.

Comparing the two spectra, there is a clear increase of Al and O peaks. The Fe- and Ni-peaks are due to a contamination when we were preparing the sample to measure EDX - they were not observed in previous measurements. We did not observe a Gd peak due to the small concentration (the EDX detection limit for Gd is approximately 0.2 at \%).

This observation reinforces the notion that the white surface is, most likely, an aluminum oxide impurity phase grown at the top of the Gd-doped SmB$_{6}$ crystals. Since this impurity could be an aluminum oxide, etching the sample should not remove it and, as a consequence, should not affect the anomalous spectra. Our result also reinterprets the reasoning behind the lack of anomalous Gd spectra in Gd-doped SmB$_{6}$ floating zone grown \cite{kunii1985} crystals: there is no impurity phase due to the growing process, hence the obtained Gd ESR spectrum is the expected one for a Gd$^{3+}$ ion in an cubic environment.

Measurements in SmB$_{6}$ Eu$^{2+}$ (x = 0.01 and 0.0004) and Er$^{3+}$ (x = 0.007) \cite{lesseux2017anharmonic} doped samples also do not show such kind of anomalous spectrum, even with the experiments being done in similar doping ranges. If the signal only appears in Gd doped samples, the signal should be related with the Gd doping. As the intensity scales well with the area of the oxide on the surface, this oxide may contain Gd and the signal could be related to Gd$^{2+}$.

Recently the synthesis of a molecular complex, containing Gd$^{2+}$ ions, was reported  \cite{completing2013macdonald}. They demonstrated that an electronic ground state 4$f^{7}$ 5$d^{1}$ for Gd$^{2+}$ exists, which goes into the direction of theoretical predictions reported by G. Wiese et al. \cite{wiese1986elektron,wiese1990possible}. They also performed ESR experiments for $T$ $\geq$ 77 K. They did not, however, observe any trace of an anomalous Gd spectrum, which is expected for this temperature range and for molecular complexes. Another recent report \cite{shallow2018kitaura} explored the origin of shallow electrons trapped in Ce-doped Gd$_{3}$Al$_{2}$Ga$_{3}$O$_{12}$. Using infrared absorption and first-principles calculations, the authors showed that defect complexes of antisite Gd$^{2+}$ ions adjacent to oxygen vacancies exist.

The above reported results point out that the Gd$^{2+}$ anomalous spectra observed is likely due to an extrinsic oxide impurity on the surface of Al-flux grown Gd:SmB$_{6}$. Such valence has been shown to be possible in oxides, especially due to defect complexes. 

\section{Conclusion}

In summary, we were able to reproduce the ESR spectra reported by G. Wiese et al. \cite{wiese1986elektron,wiese1990possible}. The ESR anomalous spectrum is highly anisotropic, is reduced for $T$ $\geq$ 6 K and is sample dependent, as was also pointed out previously \cite{wiese1990possible}. All the crystals that showed this ESR spectrum had a white region on the surface. We analyzed two pieces of the same crystal that showed the Gd anomalous spectrum. The ratio between intensities of the Gd anomalous spectra scales well with the ratio of the white region areas from the two pieces. EDX experiments showed that the white region, most likely, is due to an extrinsic Al oxide impurity phase that lies on the surface of the crystals. Finally, recent reports from literature \cite{completing2013macdonald,shallow2018kitaura} demonstrated the existence of a Gd$^{2+}$ ion in the electronic ground state 4$f^{7}$ 5$d^{1}$. This could be formed in an oxide which contains Gd and Al. Our results show that, most likely, the Gd anomalous ESR spectra in Gd-doped SmB$_{6}$, first seen by G. Wiese and confirmed by our work, may be coming from a Gd$^{2+}$ ion that lies in an extrinsic oxide phase on the surface of the crystal.

\section{Acknowledgments}

This work was supported by FAPESP\ (SP-Brazil) Grants No 2018/11364-7, 2017/10581-1, CNPq Grants No 141026/2017-0, CAPES and FINEP-Brazil. Work at Los Alamos National Laboratory was performed under the auspices of the U.S. Department of Energy, Office of Basic Energy Sciences, Division of Materials Sciences and Engineering.


\begin{thebibliography}{9}

\bibitem{dzero2016topological}
M. Dzero, X. Jing, V. Galitski, and P. Coleman
Ann. Rev. of Cond. Matt. Phys., $\textbf{7}$:249-280 (2016).

\bibitem{dzero2012theory}
M. Dzero, K. Sun, P. Coleman, and V. Galitski.
Phys. Rev. B, $\textbf{85}$(4):045130 (2012).

\bibitem{thomas2018quantum}
S.M. Thomas et al.
Phys. Rev. Lett. $\textbf{122}$(16): 166401 (2019).

\bibitem{fisk1996kondo}
Z. Fisk et al.
Phys. B $\textbf{223}$, 409-412 (1996).

\bibitem{fisk1996heavy}
Z. Fisk, J. L. Sarrao, and J. D. Thompson
Cur. Opin. in Sol. Sta. and Mat. Sci. $\textbf{1}$(1), 42-46 (1996).

\bibitem{jiao2018magnetic}
L. Jiao et al.
Sci. Adv., $\textbf{4}$(11), 4886 (2018).

\bibitem{wiese1990possible}
G. Wiese, H. Sch\"affer, and B. Elschner
Eur. Phys. Lett., $\textbf{11}$(8):791 (1990).

\bibitem{wiese1986elektron}
\textit{Elektronenspinresonanz an Gadolinium in SmB$_{6}$ - Einkristallen},
G. Wiese, Diploma Thesis, Technische Hochschule Darmstadt (1986).

\bibitem{gurin1979}
V.N. Gurin et al.
J. Less Common Met., $\textbf{67}$:115 (1979).

\bibitem{completing2013macdonald}
M.R. MacDonald et al.
J. Am. Chem. Soc., $\textbf{135}$:9857-9868 (2013).

\bibitem{shallow2018kitaura}
M. Kitaura et al.
Appl. Phys. Lett., $\textbf{113}$:041906 (2018).

\bibitem{eo2019transport}
Y-S. Eo et al.
Proc. of Nat. Acad. Sci. $\textbf{116}$(26): 12638-12641 (2019).

\bibitem{souza2018esr}
J.C. Souza et al.,
In preparation

\bibitem{kunii1985}
S. Kunii et al.
J. Magn. Magn. Mater., $\textbf{52}$:271 (1985).

\bibitem{lesseux2017anharmonic}
G.G. Lesseux et al.
AIP Adv., $\textbf{7}$(5):055709 (2017).


\end{thebibliography}
\end{document}